%% file: main_ACM.tex
\documentclass[sigconf]{Style/ACM/acmart}
\usepackage{hyperref}
\usepackage{amsmath,amsfonts}
\usepackage{graphicx}
\usepackage{xcolor}

\usepackage{booktabs}
\usepackage{array}
\usepackage{tabularx}
\usepackage{threeparttable}
\usepackage{adjustbox}
\usepackage{ragged2e}
\usepackage{makecell}
\usepackage{multirow}

\usepackage{xurl}

\setcopyright{acmlicensed}
\copyrightyear{2026}
\acmYear{2026}
\acmDOI{10.1145/3820755.3821488}
\acmConference[DocEng '26]{ACM Symposium on Document Engineering 2026}{August 25--28, 2026}{Fribourg, Switzerland}
\acmBooktitle{ACM Symposium on Document Engineering 2026 (DocEng '26), August 25--28, 2026, Fribourg, Switzerland}
\acmISBN{TBA}
\acmSubmissionID{}

\begin{document}
\title{MASCOT-Android: A Curated Dataset and Automated Collection Pipeline for Android Malware Source Code Specimens}

\input{Text/authors}

\renewcommand{\shortauthors}{Bojing Li, Duo Zhong, Prajna Bhandary, Raguvir S, Charles Maxa, Robert J Joyce, and Charles Nicholas.}

\input{Text/0_abstract_v4}

\input{Style/ACM/CCSXML}


\maketitle

\input{Text/1_introduction_v4}

\input{Text/2_Related_Work_v4}

\input{Text/3_Dataset_v4}

\input{Text/4_Automated_Source_Code_Collection_Pipeline_v4}

\input{Text/5_Code_Reuse_With_LLM_v4}

\input{Text/6_Symbol_Infor_Ablation_v4}

\input{Text/7_Discussion_v4}

\input{Text/8_Conclusion_v4}

\bibliographystyle{Style/ACM/ACM-Reference-Format}
\bibliography{references}

\end{document}

%% file: Text/authors.tex
\newcommand{\myaffiliation}{%
\affiliation{%
  \institution{University of Maryland, Baltimore County}
  \city{Baltimore}
  \state{Maryland}
  \country{USA}
}
}

\author{Bojing Li}
\myaffiliation
\email{ji8978@umbc.edu}

\author{Duo Zhong}
\myaffiliation
\email{duoz1@umbc.edu}

\author{Prajna Bhandary}
\myaffiliation
\email{prajnab1@umbc.edu}

\author{Raguvir S}
\myaffiliation
\email{nv25812@umbc.edu}

\author{Charles Maxa}
\myaffiliation
\email{cmaxa1@umbc.edu}

\author{Robert J Joyce}
\myaffiliation
\email{joyce8@umbc.edu}

\author{Charles Nicholas}
\myaffiliation
\email{nicholas@umbc.edu}
\authornote{Corresponding author.}


%% file: Text/0_abstract_v4.tex
\begin{abstract}

Compared with binaries and decompiled code, malware source code more directly reflects the attackers' original intent.
However, the scarcity of source code and the high cost of manual review make such datasets difficult to build and maintain.
We propose MASCOT-Android, a curated dataset of Android malware source code and an automated collection framework for scalable malware source code discovery on GitHub.
A key finding of our work is that repository-level documentation alone provides a strong signal for malware source code collection.
Our model extracts character-level TF-IDF features from 8,772 malware and 25,747 benign README documents and trains a LinearSVC classifier to distinguish malware repositories.
This README-only model achieves an accuracy of 96.28\% and an FPR of 1.06\% in local evaluation.
In addition, the model outputs confidence scores, allowing users to adjust the decision threshold to balance FPR and coverage, which is practical in real-world malware source code collection.

We conducted two case studies.
First, we constructed an Android malware code-reuse graph and combined it with LLM-based code detection to assess traces of LLM assistance in malware development. 
The results suggest that LLMs are already contributing, at least to some extent, to the development and propagation of malware. 
The second study performs symbolic information ablation experiments in which we gradually remove different types of symbolic information from malware source code to assess their impact on malware detection performance. 
This study shows that import statements contain highly informative signals because they are related to API usage, whereas comments and class names have limited discriminative value. 
In summary, we present a curated dataset of Android malware source code and an automated collection model, and our case studies highlight the value of source code for studying both LLM-assisted malware development and the role of symbolic information in malware detection.

\end{abstract}

%% file: Style/ACM/CCSXML.tex
\ccsdesc[500]{Security and privacy~Malware and its mitigation}
\ccsdesc[300]{Information systems~Content analysis and feature selection}
\ccsdesc[300]{Computing methodologies~Supervised learning by classification}
\ccsdesc[100]{Security and privacy~Software and application security}

%% file: Text/1_introduction_v4.tex
\section{Introduction}

The openness of the Android operating system has driven the rapid growth of its user base and mobile ecosystem, but it has also introduced a complex and evolving malware problem.
Attacks targeting mobile users worldwide in the first half of 2025 increased by 48\% more compared to the second half of 2024~\cite{kaspersky2025smartphones}. 
Message-based attacks have proliferated, and techniques for stealing one-time passwords have become widespread in real-world campaigns rather than remaining merely experimental~\cite{arntz2025android}.
Against this backdrop, systematic analyses of Android malware and the construction of research datasets have become more important than ever.

Current Android malware datasets typically provide compiled APK files together with extracted static and dynamic features to support malware detection, malware family classification, and related tasks.
However, such datasets make it difficult for researchers to understand developers' intent and malicious logic in depth.
Decompiled APK code is heavily affected by obfuscation, packing, and the compilation-decompilation process~\cite{tam2017evolution}.
Such code also lacks symbolic information, such as comments and variable names, and can distort the original control logic, making it difficult to accurately recover the developer's malicious intent.
In contrast, original malware source code preserves development logic, naming conventions, and coding style used by attackers or organizations. 
This information is valuable for understanding malicious behavior, analyzing code reuse, tracing authorship through coding style, and studying malware evolution.

Compared to malware binaries, malware source code is inherently scarce, fragmented, and ephemeral, making it difficult to collect and maintain manually. 
For example, Calleja et al. spent two years collecting only 456 specimens from underground forums and other sources~\cite{calleja2018malsource}.
The high cost of building and maintaining a malware source code dataset, therefore, requires an automated, long-term, and maintainable collection pipeline.

In this paper, we describe a well-curated, human-reviewed Android malware source code dataset, together with a reliable automated collection pipeline.
The dataset contains 1,093 specimens collected from GitHub, spanning the period from January 2011 to January 2025 and multiple loose malware categories, including backdoors, botnets, and phishing emails.
In addition, our automated collection pipeline incorporates a user-defined confidence-score threshold to filter uncertain specimens.
Even in the most challenging setting, where no confidence threshold is applied and all specimens are classified, our model achieves an average accuracy of 96.28\% and a false positive rate~(FPR) of 1.06\% across 10 repetitions of 5-fold cross-validation.

Using the proposed Android malware source code dataset, we further conduct two case studies to demonstrate its utility for future research.
RQ1: To what extent do Android malware source code specimens exhibit evidence of LLM assistance?
RQ2: How do different symbolic components of malware source code affect malware detection performance? 
The contributions of this paper are as follows:
\begin{enumerate}

\item We present a Malware Source Code Open Treasury Android~(MASCOT-Android) dataset comprising 1,093 GitHub specimens, all of which were manually reviewed, and provide curated metadata including original development dates, source URLs, and others.

\item We describe an automated pipeline for collecting malware source code specimens, which identifies malicious specimens with 96.28\% accuracy and a 1.06\% FPR. 
The pipeline also incorporates a confidence threshold to reject low confidence predictions, allowing users to trade off prediction coverage against FPR according to their needs.

\item Using this new Android malware source code dataset, we visualize overall malware evolution patterns via code-reuse relationships. 
Our analysis also reveals temporal trends in LLM-assisted malware development within the dataset.

\item We conduct ablation studies by removing different types of symbolic information from malicious source code, including imports, variable names, and others, to assess their impact on malware detection.

\end{enumerate}

The rest of this paper is organized as follows: Section~\ref{sec:related_work} presents previous research on Android dataset collection. 
Section~\ref{sec:dataset} building the MASCOT-Android dataset and statistical information.
Section~\ref{sec:automated_malware_framework} describes the structure of our automated malware collection pipeline and its collection performance in local tests and real-world tasks. 
Section~\ref{sec:code_reuse_LLM} examines the evolution of LLM-assisted malware source code development.
Section~\ref{sec:ablation_experiments} presents the ablation experiments for symbolic information with regard to malware detection.
Section~\ref{sec:discussion}, and Section~\ref{sec:conclusion} discuss our findings and conclusions.

%% file: Text/2_Related_Work_v4.tex
\section{Related Work}
\label{sec:related_work}

Existing datasets typically emphasize either large-scale specimen collection or comprehensive static and dynamic features extracted from APK files.
The Canadian Institute for Cybersecurity has introduced a series of Android malware datasets.
In CICAndMal2017, 429 Android malware specimens were executed in a controlled environment, and network traffic was captured at three stages: before installation, before reboot, and after reboot~\cite{lashkari2018toward}.
CCCS-CIC-AndMal-2020 expanded this work to 200,000 benign specimens and 200,000 Android malware specimens~\cite{keyes2021entroplyzer}.
Their feature set includes both static features, such as permissions and services, and dynamic features, such as system calls and network traffic.
MH-1M is a recent large-scale Android malware dataset with a comprehensive feature set~\cite{bragancca2025mh}.
It comprises 1,340,515 Android APK files and 22,810 extracted features, and is intended to support future research through its rich metadata and broad coverage.

However, few studies have focused on collecting Android specimens in their original source code form, which can indicate the author's intent.
Malsource and MASCOT are representative malware source code datasets, but both are limited to the Windows platform~\cite{calleja2018malsource,11401016}.
For Android, to the best of our knowledge, the publicly released dataset by Github user \texttt{d-Raco} contains source code of only 97 specimens and was described as the largest dataset available in 2022~\cite{draco_android_malware_source_code_analysis}. 
The scarcity of Android malware source code datasets stems from the lack of a reliable pipeline for collecting trustworthy malware source code.
As far as we know, SourceFinder is the only malware source code dataset constructed using an automated collection pipeline~\cite{rokon2020sourcefinder}.
Those authors manually labeled 1,000 specimens for training, and the resulting classification performance suggests that reliable automated malware source code collection remains challenging.

In summary, Android malware source code remains scarce, and prior efforts to collect such code automatically have achieved only limited success.

%% file: Text/3_Dataset_v4.tex
\section{Dataset}
\label{sec:dataset}

Following Li et al., we built the MASCOT-Android dataset through keyword search, author-based snowballing, fork-based snowballing, and iterative repetition of these steps~\cite{11401016}.
This new dataset contains 1,093 Android malware source code specimens collected from GitHub and spans the period from January 2011 to January 2025.
Each specimen is accompanied by metadata, including the GitHub repository name, the development date, the keywords or clues used to locate the specimen, the URL, and additional relevant information.
To ensure the quality of the MASCOT-Android dataset, we also followed the manual review procedure of Li et al.~\cite{11401016}. 
We first reviewed the README files to determine whether each specimen claimed to be malicious or benign.
For forked specimens, we used the GitHub comparison function to examine source code modifications and retained only those with substantive functional updates, rather than changes limited to README files, author information, or parameter settings.
Following the same collection and review procedure as MASCOT, MASCOT-Android provides a consistent basis for cross-platform comparison.

Because of this methodological consistency, we can compare MASCOT and MASCOT-Android to assess differences between Windows and Android malware while reducing bias from the collection procedure.
Table~\ref{table:windows_android_source_compare} shows the sources of the specimens in these two datasets.
In the Windows dataset, author-related clues account for a large share of the data, suggesting that the same developers often produce multiple malware specimens.
In contrast, for the Android dataset, fork-related clues account for a larger share, suggesting more frequent code reuse from existing malware and a higher prevalence of derived variants.
In addition, compared with the Windows dataset, fewer Android specimens were identified through dataset-related clues, suggesting that publicly shared collections of Android malware source code are less common and that such code may be more difficult to retrieve from public sources.

\begin{figure}[t]
    \centering
    \includegraphics[width=\linewidth]{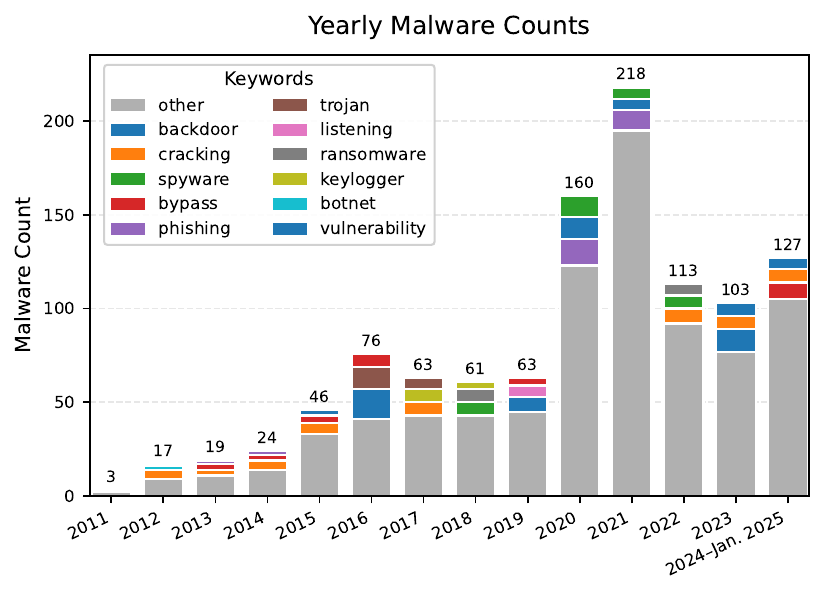}
    \caption{Yearly distribution of Android malware source code specimens by retrieval keyword.}
    \label{fig:yearly_android_malware_keyword_distribution}
\end{figure}

Figure~\ref{fig:yearly_android_malware_keyword_distribution} and Table~\ref{table:windows_android_keyword_compare} present the top keywords used to search for malware specimens on the Windows and Android platforms under the same collection method.
The results show that malware specimens across platforms exhibit distinct keyword distributions, indicating a strong association between keyword selection and platform-specific malware characteristics.
For example, the most frequently used keyword in Windows, \texttt{keylogger}, is used by attackers to record what the victim types on the keyboard.
In contrast, the keywords \texttt{crack} and \texttt{bypass} are commonly used in Android malware searches because it is often associated with attempts to bypass or crack the screen lock on Android mobile devices.
Therefore, when searching for malware specimens across different platforms, it is necessary to use keywords that are closely aligned with the platform's characteristics.

\input{Tables/3_dataset_source_compare}
\input{Tables/3_dataset_top_keywords}

%% file: Tables/3_dataset_source_compare.tex
\begin{table}[t]
\centering
\begin{threeparttable}
\caption{Comparison of collection sources for MASCOT and MASCOT-Android.}
\label{table:windows_android_source_compare}
\footnotesize
\setlength{\tabcolsep}{3pt}
\begin{tabular*}{\columnwidth}{@{\extracolsep{\fill}} l l l c c c}
\toprule
\textbf{Dataset} & \textbf{Platform} & \textbf{Fork} & \textbf{Existing Sets} & \textbf{Author} & \textbf{Keywords} \\
\midrule
MASCOT & Windows & 12.2\% & 21.3\% & 25.1\% & 41.4\% \\
MASCOT-Android & Android & 28.4\% & 8.9\%  & 5.9\%  & 56.8\% \\
\bottomrule
\end{tabular*}
\begin{tablenotes}[flushleft]
\footnotesize
\item Note: Existing Sets refers to specimens collected from existing small malware source code collections hosted on GitHub.
\end{tablenotes}
\end{threeparttable}
\end{table}


%% file: Tables/3_dataset_top_keywords.tex
\begin{table}[t]
\centering
\begin{threeparttable}
\caption{Comparison of top-5 keyword clues in MASCOT and MASCOT-Android.}
\label{table:windows_android_keyword_compare}
\footnotesize
\setlength{\tabcolsep}{3pt}
\renewcommand{\arraystretch}{1.05}
\begin{tabular*}{\columnwidth}{@{\extracolsep{\fill}} l c c c c c}
\toprule
\textbf{Dataset} 
& \textbf{\#1} 
& \textbf{\#2} 
& \textbf{\#3} 
& \textbf{\#4} 
& \textbf{\#5} \\
\midrule

MASCOT 
& \makecell[c]{keylogger\\12.3\%}
& \makecell[c]{trojan\\5.3\%}
& \makecell[c]{reverse shell\\3.3\%}
& \makecell[c]{virus\\2.4\%}
& \makecell[c]{hack\\1.6\%} \\

\addlinespace[3pt]

\makecell[l]{MASCOT-\\Android}
& \makecell[c]{crack\\6.3\%}
& \makecell[c]{backdoor\\6.3\%}
& \makecell[c]{spyware\\4.8\%}
& \makecell[c]{bypass\\4.8\%}
& \makecell[c]{phish\\4.4\%} \\

\bottomrule
\end{tabular*}
\end{threeparttable}
\end{table}

%% file: Text/4_Automated_Source_Code_Collection_Pipeline_v4.tex
\section{Automated Malware Source Code Collection Pipeline}
\label{sec:automated_malware_framework}

Although malware source code datasets have several advantages, including being unobfuscated, preserving symbolic information, and reflecting real malicious logic, their main limitation lies in the difficulty of collection.
Calleja et al. ~\cite{calleja2018malsource} spent two years collecting and reviewing 456 malware source code specimens, while Li et al.~\cite{11401016} spent one year collecting and reviewing 6,032 from GitHub.
Rokon et al. manually reviewed 1,000 malware source code specimens~\cite{rokon2020sourcefinder} but did not report the time required.
In this study, we spent approximately five to six months collecting and reviewing 1,093 Android malware source code specimens from GitHub.
To the best of our knowledge, these datasets constitute the largest publicly available curated malware source code datasets.

Given the difficulty of manual collection and review, we argue that although manual review can improve dataset quality, it does not scale enough to support future updates and long-term maintenance.
Nevertheless, these manually reviewed specimens, together with the experience gained from constructing the datasets discussed above, provide a foundation for developing an automated pipeline for collecting malware source code.

\subsection{Sources of Malware Source code Specimens}

Previous studies collected malware source code from public platforms and underground forums~\cite{calleja2018malsource}.
Underground forums are less suitable for automated collection because they often lack complete development metadata and are hard for researchers to access.

Compared with other sources, we consider GitHub most suitable for automated malware source code collection for the following reasons:
(a)~Previous research has shown that many malware specimens are hosted on GitHub~\cite{rokon2020sourcefinder}. 
GitHub rarely blocks repositories solely because of malicious keywords, as we also observed during the collection of the MASCOT-Android dataset and shown in Table~\ref{table:windows_android_keyword_compare}. 
(b)~GitHub repositories provide development histories and rich metadata, which are valuable for malware analysis. 
For example, fork relationships provide useful evidence for analyzing malware evolution.
(c)~As one of the largest code hosting platforms, GitHub provides a mature token and API system that supports large-scale repository search, retrieval, and analysis, which facilitates automated malware source code collection. 
(d)~As we will explain later, README files in GitHub provide useful evidence to distinguish malicious repositories from benign ones.

\begin{figure}[t]
    \centering
    \includegraphics[width=\linewidth]{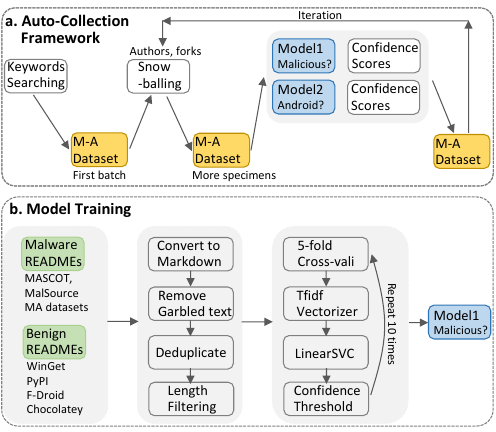}
    \caption{Overall architecture of our automated malware source code collection framework and model training}
    \label{fig:overall_architecture}
\end{figure}

Once the data is collected, we apply our automated malware source code collection pipeline, as shown in Figure~\ref{fig:overall_architecture}.
This pipeline extends the manual review process used to construct the MASCOT-Android dataset and retains its main steps, including keyword search, snowballing based on forks and authors, and iterative expansion.
The key difference is that manual review is replaced by two LinearSVC models: one to distinguish malicious specimens from benign ones, and another to distinguish Android specimens from non-Android specimens.

\subsection{Model Training}

To train these models, we first collect README files from MalSource~\cite{calleja2018malsource}, MASCOT~\cite{11401016}, and our MASCOT-Android dataset, all of which have been manually reviewed and considered as the ground truth.
MalSource and MASCOT provide malware README files for Windows, while the MASCOT-Android dataset comprises Android malware.

Second, we collect benign README files from secure platforms with security and malware-prevention mechanisms, including Winget, PyPI, Chocolatey, and F-Droid. 
We assume that popular software on these platforms is unlikely to explicitly describe malicious behavior in their README files.

Third, we detect and remove garbled README files. We convert all benign or malware files from sources above into a uniform Markdown format.
During this conversion process, some files may contain garbled text due to character encoding or compatibility issues.
We use charset\_normalizer~\cite{charset_normalizer} to identify documents that may contain garbled text and manually remove them.
Approximately 60 out of 5,000 README files contain garbled text and are removed.

The fourth step is deduplication using MD5 hashes.
Forked repositories usually focus more on functional updates, but may contain identical README files. Finally, we apply length filtering by removing README files shorter than 40 bytes.

\input{Tables/4_README_dataset_source}

As shown in Table~\ref{tab:readme_dataset_composition}, after applying all preprocessing steps, we obtained 25,747 benign and 8,772 malicious README files.
Notably, in some source datasets, the number of retained README files exceeds the number of app specimens.
This occurs because complex modern Android applications often contain multiple submodules, and developers may provide a separate README file for each submodule.
As a result, a single Android project may contain multiple README files, sometimes between 20 and 40.

After preprocessing, we use TfidfVectorizer from scikit-learn to tokenize the processed source code and construct feature vocabularies~\cite{tfidfvectorizer}.
The vectorizer operates at the character level, uses 2 to 10 character n-grams, and sets \texttt{ min\_df=0.05} and \texttt{max\_df=0.95} to keep only features that appear in 5\% to 95\% of the specimens.
This removes overly common features with limited discriminative value and overly rare features that are more likely to capture noise or specimen-specific artifacts rather than generalizable classification signals. 
Other parameters remain at their default values.

\begin{figure}[t]
    \centering
    \includegraphics[width=\linewidth]{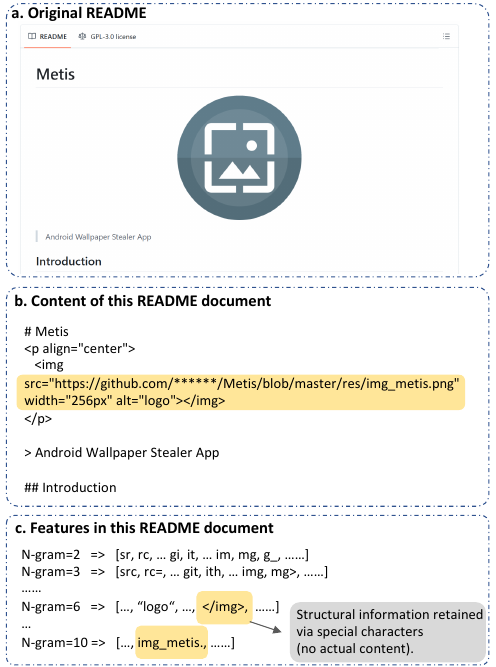}
    \caption{Processed README files and preserved structure features}
    \label{fig:readme_features}
\end{figure}

This feature-processing method enables our pipeline to retain structural information encoded in special characters, as shown in Figure~\ref{fig:readme_features}. 
We believe these formatting elements, such as images, hyperlinks, font styles, and emojis, may provide useful signals for distinguishing benign from malicious specimens.

\subsection{Local Evaluation}

Using the constructed feature vocabularies, we employ LinearSVC from scikit-learn for malware detection~\cite{linearsvc}.
Since the vocabulary depends on the training split in five-fold cross-validation, we repeat the process with 10 random seeds to reduce bias introduced by a particular training-testing split.

As shown in Table~\ref{tab:model_performance_seeds}, Model 1 achieves an overall average accuracy of 96.28\%, a Macro-F1 score of 94.95\%, and an FPR of 1.06\% for classifying README files as malicious or benign. In comparison, Model 2 achieves lower performance (86.75\% accuracy, 78.88\% Macro-F1, 5.54\% FPR) for platform classification.
This difference is likely due to the cross-platform nature of many malware samples. 
For example, some specimens include both Android components and Linux scripts, making it difficult to assign a single platform label based solely on README content.
All results above are obtained without applying confidence thresholds to reject low-confidence cases. Therefore, the prediction coverage is 100\% here.

\begin{figure}[t]
    \centering
    \includegraphics[width=\linewidth]{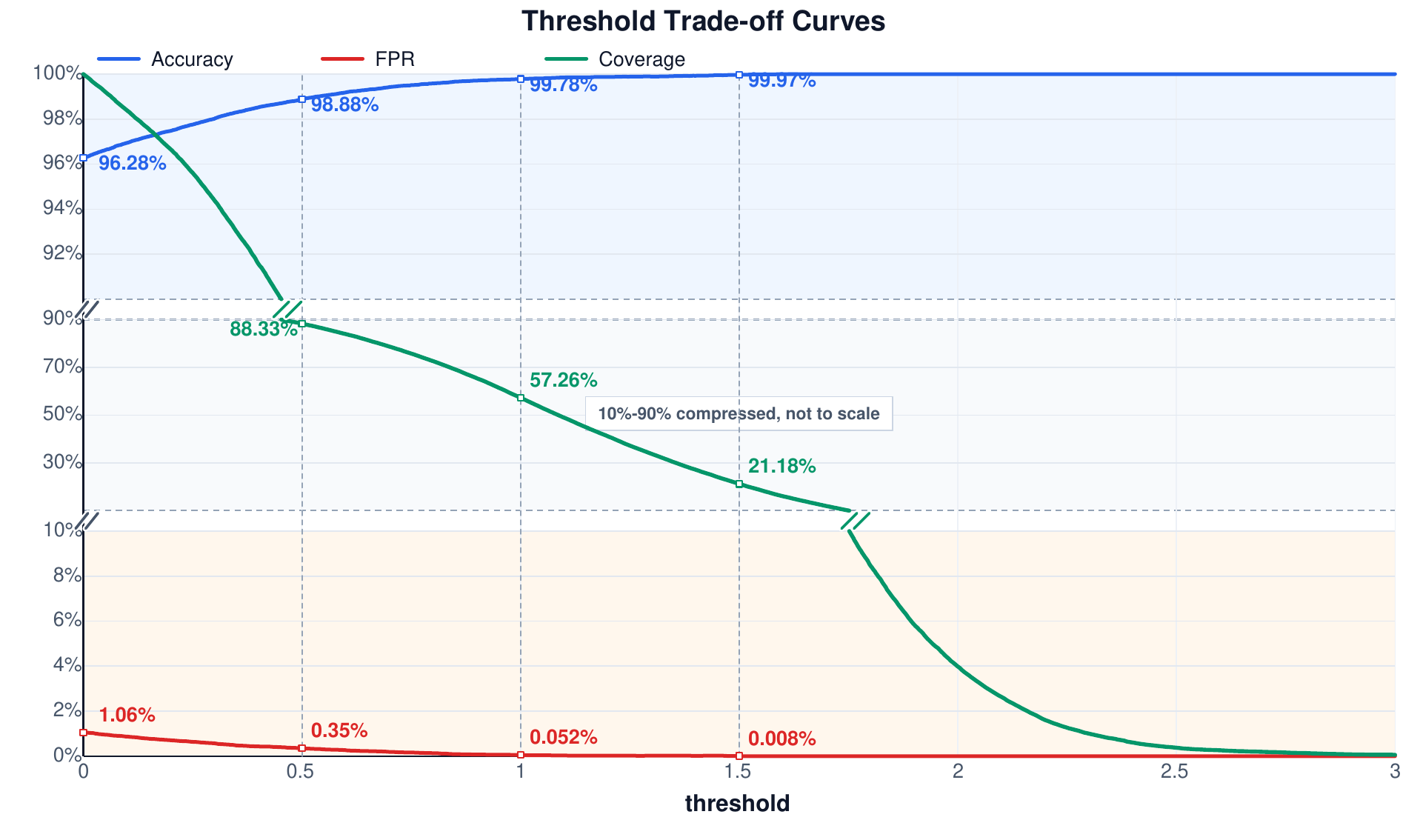}
    \caption{Confidence-threshold trade-off among FPR, accuracy, and coverage.}
    \label{fig:threshold_tradeoff}
\end{figure}

Applying thresholds can further improve accuracy and reduce FPR, at the cost of reduced coverage.
As shown in Figure~\ref{fig:threshold_tradeoff}, increasing the confidence thresholds improves accuracy and reduces FPR but decreases coverage. 
When the threshold increases from 0 to 0.5, accuracy improves from 96.28\% to 98.88\%, and FPR drops from 1.06\% to 0.35\%, while only 11.67\% of specimens are rejected. 
Increasing the threshold further to 1.0 improves accuracy to 99.78\% and reduces FPR to 0.052\%, but at the cost of rejecting 42.74\% of specimens.
This demonstrates a clear trade-off between performance and prediction coverage. 
Note that all test specimens were preprocessed to exclude garbled, extremely short, or otherwise low-quality text, whereas GitHub README files may not meet the same quality standards.
Therefore, from a practical perspective, a threshold between 0.5 and 1.0 provides a reasonable balance, especially given the variability in real-world README files.

\input{Tables/4_classification_performance}

\subsection{Real-World Evaluation}

Due to limitations of the GitHub API, we evaluate our pipeline on GitHub data collected over weekly intervals. 
As a case study, we use Android malware source code specimens whose latest push dates fall between March 2, 2025, and March 8, 2025.
Our pipeline required 578 minutes to perform keyword searches and multiple rounds of snowballing, identifying 11,982 suspicious repositories on GitHub.
After deduplication and the removal of specimens without a README file, 595 specimens remained, of which 83 were classified as Android malware, and the remaining specimens were classified as Windows malware.
The numbers reported above were obtained before applying the confidence score threshold, and the corresponding results were saved in a CSV file.
This design allows us to retain as many candidate specimens as possible and then flexibly adjust the confidence threshold and filtering criteria according to different experimental needs.

To assess the quality of the collected Android malware candidates, we manually reviewed all 83 specimens classified as Android malware.
We found that 9 of the 83 specimens were actually benign, and 2 of these 9 specimens were also misclassified on the platform.
In this case, we observed 9 false positives among the 83 repositories classified as Android malware. The number of misclassified specimens is higher than expected compared to the local experiments.
We attribute this discrepancy to the complexity of README files in real-world collection tasks.
For example, two of the misclassified README files were written in Korean and Portuguese, respectively.
Since our discriminative model was not explicitly optimized for multilingual processing, its performance on multilingual inputs remains limited in both accuracy and confidence scores.
Three other misclassified files contained only one sentence or even only a few words.
Such brief files were filtered out in our local tests, but they cannot be excluded in real-world collection tasks.
The remaining misclassified specimens contained substantial structural information, such as images, but little textual content.

Nevertheless, the confidence score threshold was effective in this case.
We found that although there were more misclassified specimens than in the local tests, all of them had low confidence scores, with the highest score being 0.38.
When an appropriate confidence score threshold between 0.5 and 1.0 is applied, as suggested above, all misclassified specimens in this case would be filtered out and excluded from the final malware source code dataset.

In summary, these results indicate that our automated malware source code collection pipeline is practical and effective. 
In a one-week real-world collection task, we observed that the model did not maintain the low FPR of 1.06\% observed in local tests when facing complex inputs, such as multilingual README files, short textual descriptions, and structurally rich but text-sparse repositories.
However, by applying a confidence score threshold, the framework can still filter out specimens with uncertain predictions and maintain a malware source code dataset of high quality, although this comes at the cost of coverage.

%% file: Tables/4_README_dataset_source.tex
\begin{table}[t]
\centering
\caption{Composition of the README dataset by label and platform.}
\label{tab:readme_dataset_composition}
\begin{tabular}{p{0.16\linewidth} p{0.33\linewidth} p{0.25\linewidth} r}
\toprule
\textbf{Label} & \textbf{Windows} & \textbf{Android} & \textbf{Total} \\
\midrule
Benign &
WinGet: 4,220 \newline
PyPI: 10,079 \newline
Chocolatey: 1,497 
&
F-Droid: 9,951 
&
\textbf{25,747} \\
\midrule
Malicious &
MASCOT: 6,789 \newline
MalSource: 75 
&
MASCOT-\newline Android: 1,908 
&
\textbf{8,772} \\
\midrule
\textbf{Total} &
\textbf{22,660} &
\textbf{11,859} &
\textbf{34,519} \\
\bottomrule
\end{tabular}
\end{table}

%% file: Tables/4_classification_performance.tex
\begin{table*}[t]
\centering
\caption{Model performance across ten random seeds.}
\label{tab:model_performance_seeds}
\scriptsize
\setlength{\tabcolsep}{3.5pt}
\renewcommand{\arraystretch}{1.15}

\begin{adjustbox}{max width=\textwidth}
\begin{threeparttable}
\begin{tabular}{llccccccccccc}
\toprule
\textbf{Model} & \textbf{Metric} 
& \textbf{Seed1} & \textbf{Seed2} & \textbf{Seed3} & \textbf{Seed4} & \textbf{Seed5} 
& \textbf{Seed6} & \textbf{Seed7} & \textbf{Seed8} & \textbf{Seed9} & \textbf{Seed10} & \textbf{Avg.} \\
\midrule

\multirow{4}{*}{\makecell[l]{Malicious/\\Benign}}
& Acc     & 96.54\% & 96.09\% & 96.02\% & 96.35\% & 96.41\% & 96.65\% & 96.09\% & 96.39\% & 96.16\% & 96.15\% &\textbf{96.28\%} \\
& F1-macro      & 95.32\% & 94.67\% & 95.58\% & 95.05\% & 95.15\% & 95.47\% & 94.66\% & 95.10\% & 94.79\% & 94.76\% &\textbf{94.95\%} \\
& FPR     & 1.09\%  & 1.05\%  & 1.09\%  & 1.09\%  & 1.24\%  & 0.93\%  & 0.89\%  & 0.99\%  & 1.11\%  & 1.09\%  &\textbf{1.06\%} \\
& Feature & 38.06m   & 38.41m   & 38.04m   & 37.85m   & 38.15m   & 37.87m   & 37.79m   & 38.00m   & 38.01m   & 37.85m   &\textbf{38.00m} \\
\midrule

\multirow{4}{*}{\makecell[l]{Android/\\Windows}}
& Acc     & 86.21\% & 86.89\% & 86.84\% & 87.18\% & 86.15\% & 87.75\% & 87.12\% & 85.87\% & 87.69\% & 85.81\% &\textbf{86.75\%} \\
& F1-macro      & 78.08\% & 79.35\% & 79.37\% & 79.64\% & 77.61\% & 80.71\% & 79.44\% & 77.53\% & 79.70\% & 77.37\% &\textbf{78.88\%} \\
& FPR     & 5.97\%  & 5.83\%  & 6.05\%  & 5.39\%  & 5.46\%  & 5.32\%  & 5.24\%  & 6.19\%  & 3.86\%  & 6.12\%  &\textbf{5.54\%} \\
& Feature & 10.10m  & 10.33m  & 10.07m  & 9.86m   & 10.03m  & 9.60m   & 9.91m   & 10.18m  & 10.26m  & 9.94m   &\textbf{10.03m} \\

\bottomrule
\end{tabular}

\begin{tablenotes}[flushleft]
\footnotesize
\item[] \textit{Note:} Feature denotes the number of extracted features, where `m' represents millions. Avg. reports the metric computed over the pooled predictions from all repeated cross-validation runs, rather than the arithmetic mean.
\end{tablenotes}

\end{threeparttable}
\end{adjustbox}
\vspace{-1.5em}

\end{table*}

%% file: Text/5_Code_Reuse_With_LLM_v4.tex
\section{Code Reuse Network with LLM Annotation}
\label{sec:code_reuse_LLM}

In this section, we examine the analytical value of malware source code by asking RQ1: To what extent do Android malware source code specimens exhibit evidence of LLM assistance?

This question is especially important because malware development is often characterized by frequent code reuse. 
Calleja et al. noted that attackers often adapt functionalities from existing malware when developing new specimens~\cite{calleja2018malsource}.
This pattern is also evident in the MASCOT and MASCOT-Android datasets, as shown in Table~\ref{table:windows_android_source_compare}, where 12.2\% and 28.4\% of the specimens, respectively, were collected from forks created through code reuse.
Although major LLM companies have adopted multiple safeguards to prevent the generation of malicious code~\cite{openai_system_card}, LLMs remain vulnerable to misuse.
For example, Chao et al. showed that semantic jailbreaks can be achieved even with black-box access to an LLM~\cite{chao2025jailbreaking}.
Therefore, if LLM-assisted code is already present in malware development, frequent code reuse may further propagate it across malware specimens.

\begin{figure}[b]
    \centering
    \includegraphics[width=\linewidth]{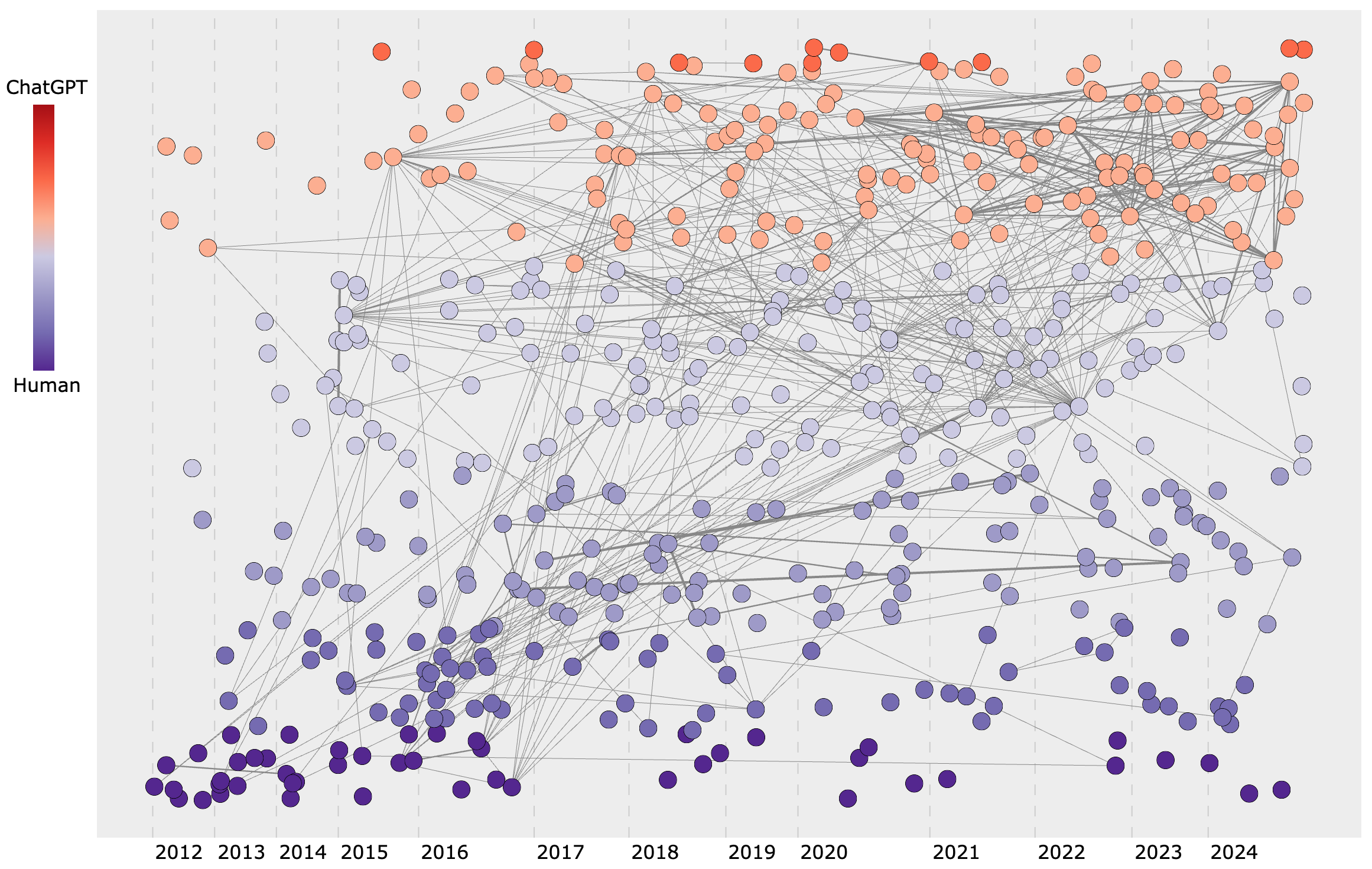}
    \caption{LLM-annotated code reuse network of Android malware source code specimens~(Java specimens only).}
    \label{fig:visualization}
\end{figure}

We use the MASCOT-Android dataset to construct a code reuse graph and annotate it with LLM trace signals to investigate this question.
First, we use NiCad, a hybrid code clone-detection tool, to identify code reuse~\cite{cordy2011nicad}.
NiCad parses source code at a specified granularity and converts it to comparable text by removing comments and normalizing the formatting.
This tool also normalizes the code by standardizing identifiers and selected syntactic forms.
It identifies code clones by comparing the longest common subsequence across standardized code snippets.

After identifying code reuse, we combine GPTZero~\cite{adam2026gptzero} and GPTSniffer~\cite{nguyen2024gptsniffer} to detect LLM traces in malware source code.
GPTZero uses a hierarchical classification scheme that distinguishes among pure AI, AI-polished, and AI-paraphrased content after detecting LLM traces.
Since GPTZero is primarily a general-purpose AI text detector, we also introduce GPTSniffer as a complementary code-oriented detector.
GPTSniffer is an AI code detector built on the CodeBERT model~\cite{feng2020codebert}.
As CodeBERT is pretrained on large-scale source code corpora, GPTSniffer has shown strong performance in detecting LLM traces in source code, but the overall performance drops when evaluating from unknown sources~\cite{nguyen2024gptsniffer}.

\begin{figure*}[t]
    \centering
    \includegraphics[width=\textwidth]{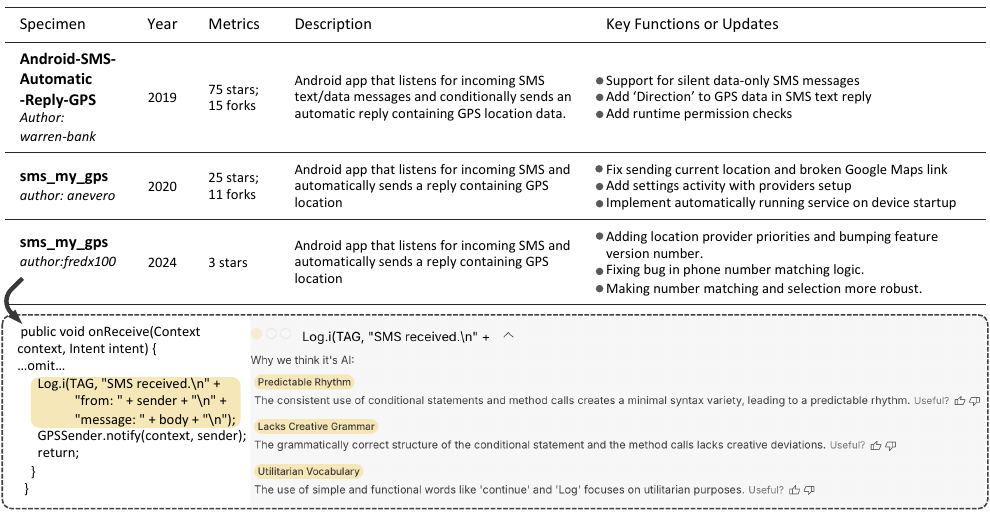}
    \caption{Code reuse lineage of an Android malware, with LLM traces observed in the 2024 variant.}
    \label{fig:code_reuse_LLM_traces}
\end{figure*}

Based on code-reuse detection and LLM trace annotations from GPTZero and GPTSniffer, we visualize the resulting network in Figure~\ref{fig:visualization}.
Each edge represents code reuse between two specimens, and node color indicates the intensity of LLM trace signals.
Figure~\ref{fig:visualization} suggests three main observations.
First, we observed that the number of specimens labeled as LLM-assisted began to increase between 2017 and 2018, and that these specimens formed more densely connected regions after 2022.
This temporal trend is broadly consistent with several milestones in language model development, including the introduction of the Transformer architecture in 2017~\cite{vaswani2017attention}, the release of BERT~\cite{devlin2019bert} and GPT-1~\cite{Radford2018ImprovingLU} in 2018, and the release of ChatGPT in 2022.
This pattern suggests that LLM-assisted traces in Android malware source code may be associated with the broader development and availability of LLMs.
Second, according to GPTZero and GPTSniffer, specimens with very strong LLM-assisted signals~(depicted in red) are uncommon, whereas a substantial number of Android malware specimens exhibit moderate LLM-assisted signals.
This pattern suggests that LLMs are currently more likely to assist with incremental development and refinement than to generate entire malware projects from scratch.
Third, although we combined a general-purpose detector and a code-oriented detector, some specimens before 2018 are still misclassified as LLM-assisted.
These false positives highlight the need for more up-to-date detection methods tailored for source code.

To assess whether LLMs may contribute to the spread of malicious code via code-reuse networks, we use the \texttt{SMSReceiver} lineage as a case study.
As shown in Figure~\ref{fig:code_reuse_LLM_traces}, GitHub user \texttt{warren-bank} created the project in 2019 and publicly released an initial version in 2020~\cite{warrenbank_android_sms}.
The initial project aims to track victims' locations through specially formatted text messages.
In 2021, another user, \texttt{anevero}, extended the project with additional functionality, including automatic execution at device startup, and released the updated version~\cite{anevero_sms_my_gps}.
In 2024, \texttt{fredx100} further expanded the project by allowing attackers to select the GPS provider, thus obtaining more accurate victim locations~\cite{fredx100_sms_my_gps_2024}.
In the 2024 version, one of the key malicious components is \texttt{SMSReceiver.java}, which parses text messages, writes logs, and invokes GPS-related functions.
In \texttt{SMSReceiver.java}, GPTSniffer assigns an LLM generation probability of $p=0.9990$. GPTZero 4.4b classifies the file as mainly human-written but still identifies 13 LLM traces, suggesting that it may have been revised or polished with LLM assistance, as shown in Figure~\ref{fig:code_reuse_LLM_traces}.

The evolution of this lineage reflects a common pattern in malware development: once a representative malware specimen is released, attackers may reuse and extend it to incorporate more complex functionality.
In this case, the trajectory also appears to exhibit signs of LLM assistance, as highlighted in Figure~\ref{fig:code_reuse_LLM_traces}, which is consistent with the broader pattern shown in Figure~\ref{fig:visualization}.
Taken together, these findings suggest that LLM-like code signals are present in recent Android malware source code and may appear along code-reuse lineages.

%% file: Text/6_Symbol_Infor_Ablation_v4.tex
\section{Ablation Experiments for Symbol Information}
\label{sec:ablation_experiments}

\begin{figure}[t]
    \centering
    \includegraphics[width=\linewidth]{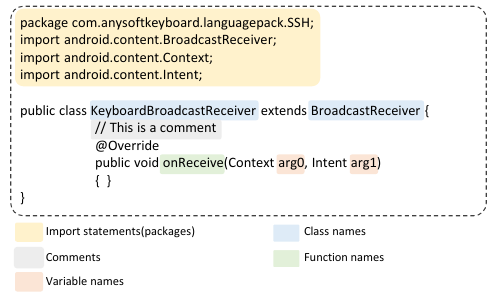}
    \caption{Different symbolic information in JAVA code}
    \label{fig:symbolic_information}
\end{figure}

As shown in Figure~\ref{fig:symbolic_information}, symbolic information can reveal the semantics of program components and, to some extent, how these components are used, making it valuable for multiple cybersecurity tasks.
Someya et al. used semantic information, such as API names parsed by IDA Pro, together with function call graphs and attention mechanisms to build an interpretable malware classification method~\cite{someya2023fcgat}.
Cozzi et al. used symbolic information to distinguish malware variants and build an IoT malware genealogy~\cite{cozzi2020tangled}.
Yakdan et al. proposed the DREAM++ decompiler, which renames variables according to contextual semantics and significantly improves the readability of decompiled code~\cite{7546501}.
These studies show that symbolic information is useful for malware classification interpretability, malware evolution analysis, and decompiled code readability.
However, the role of different symbolic components of source code in malware detection has not been systematically and quantitatively compared. 
Based on this gap, we focus on another potential value of malware source code. RQ2: How do different symbolic components of malware source code affect malware detection performance?

To answer this question, we selected 606 Java-based Android malware source code specimens from the MASCOT-Android dataset and 606 benign specimens from F-Droid~\cite{fdroid_about}.
We then removed various symbolic components from the source code and conducted an ablation study to compare malware detection performance across configurations.
As shown in Table~\ref{tab:ablation_performance}, configurations c1 to c5 progressively remove import statements, comments, class names, function names, and variable names to assess the impact of different symbolic components on malware detection.
In addition, we designed configuration c6 to approximate a decompiled-code-like setting.
In c6, we retain import statements but remove all other symbolic information.
This setting is motivated by the fact that library dependencies and API usage are often easier to recover from decompiled code, whereas fine-grained source-level information, such as original class names, function names, variable names, and comments, is more likely to be lost, removed, or rewritten during compilation, obfuscation, and decompilation.

This experiment consists of the following steps:
We use tree-sitter~\cite{tree_sitter_docs} to parse the Java source code and identify different components, including import statements, comments, class names, function names, and variable names.
We then normalize or remove the corresponding symbolic information under different experimental configurations to simulate the gradual loss of source-level symbolic information.
For example, we rename the original variable name to \texttt{variable1} and remove comments.
Finally, we combine all processed .java files into a single .txt file as the input representation for each specimen.

We follow TfidfVectorizer~\cite{tfidfvectorizer} and LinearSVC~\cite{linearsvc} processes described in Section~\ref{sec:automated_malware_framework} to construct character-level three to six n-gram vocabularies and perform malware detection. 
We also apply 5-fold cross-validation, repeated 10 random seeds, to reduce bias introduced by data splitting.

\input{Tables/6_ablation_result}

Table~\ref{tab:ablation_performance} shows the performance of malware detection under different source code configurations.
Overall, the results show that different symbolic components contribute differently to malware detection.

We have three main observations. 

(1) Compared with the original configuration c0, configuration c1 removed import statements and shows a drop in malware detection accuracy despite only a small reduction(1.1\%) in feature vocabulary size. 
This result suggests that, compared to other source-level symbolic components, library dependencies, package information, and API-related information reflected by import statements provide strong discriminative signals for malware detection.
This observation is also consistent with prior research on malware detection, which often treats library dependencies and API usage as important features~\cite{sami2010malware}.

(2) Comments, class names, function names, and variable names do not provide additional discriminative information.
Our results show that, after removing comments and normalizing variable names, the average number of feature vocabulary decreases by approximately 20.9\% and 30.1\%, respectively.
In contrast, the accuracy of malware detection increases slightly. 
We further adjusted several experimental settings, including the ratio of malicious to benign specimens, the n-gram parameters, and the tokenizer granularity~(char or word). 
The results remained consistent across these settings.
This suggests that these symbolic components may be useful for malware evolution analysis, interpretable security analysis, or decompiled code readability.
But they do not provide highly discriminative signals for malware detection from our current experimental setup. 
One possible explanation is that these components may be similar to their counterparts in benign code and, therefore, may not provide discriminative information.

(3)The decompile-like configuration c6 achieves the best performance while using a much smaller feature vocabulary, as shown in Table~\ref{tab:ablation_performance}.
This result suggests that if decompilation accurately preserves import statements and the main logic and structure of the code body, decompiled code may still contain sufficient information for high-precision malware detection.

\input{Tables/6_top_features_in_ablation}

To further explain these observations, we extract the supportive character n-gram features from the classification model and organize these features into groups, as shown in Table~\ref{tab:top_features_in_ablation}.
For the malicious class, \texttt{sms} in different uppercase and lowercase combinations is among the strongest features driving the model toward this class.
Certain character n-grams derived from words, such as \texttt{trojan} and \texttt{phishing}, also contribute strongly to the malicious class.
For the benign class, the strongest feature groups are less directly security-related. 
They include documentation or formatting patterns, Android resource references, and UI/application development terms.
For example, the \texttt{(space)*(space)} pattern reflects a common formatting pattern in Java block comments, while n-grams derived from terms such as \texttt{color} and \texttt{View} are related to ordinary UI/application development.

In summary, this section evaluates another potential research value of malware source code: assessing how different symbolic components of source code support malware detection.

%% file: Tables/6_ablation_result.tex
\begin{table}[t]
\centering
\caption{Average accuracy, vocabulary size, and runtime under different preprocessing configurations.}
\label{tab:ablation_performance}
\begin{threeparttable}
\footnotesize
\begin{tabular}{lccc}
\toprule
\textbf{Configuration} & \textbf{Accuracy (avg.)} & \textbf{Vocabulary (avg.)} & \textbf{Time (avg.)} \\
\midrule
c0  & 89.79\% & 5,180,693 & 391s \\
c1            & 88.86\% & 5,123,673 & 369s \\
c2            & 88.94\% & 4,052,298 & 281s \\
c3            & 89.02\% & 4,027,121 & 255s \\
c4            & 89.07\% & 3,945,530 & 244s \\
c5            & 89.50\% & 2,756,901 & 157s \\
c6            & 90.62\% & 3,061,125 & 340s \\
\bottomrule
\end{tabular}
\begin{tablenotes}[flushleft]
\footnotesize
\item c0: Original source code
\item c1: w/o imports
\item c2: w/o imports and comments
\item c3: w/o imports, comments, and class names
\item c4: w/o imports, comments, class names, and func names
\item c5: w/o imports, comments, class names, func names, and variable names
\item c6: Decompiled-code-like representation with only imports retained
\end{tablenotes}
\end{threeparttable}
\end{table}

%% file: Tables/6_top_features_in_ablation.tex
\begin{table}[t]
\centering
\caption{Representative high-weight character n-gram feature groups.}
\label{tab:top_features_in_ablation}
\footnotesize
\setlength{\tabcolsep}{4pt}
\renewcommand{\arraystretch}{1.1}

\begin{tabular}{p{0.18\linewidth} p{0.30\linewidth} p{0.42\linewidth}}
\toprule
\textbf{Class} & \textbf{Feature group} & \textbf{Representative n-grams} \\
\midrule

\multirow{4}{*}[-6ex]{Malicious}
& \textbf{SMS-related terms}
& \texttt{sms}, \texttt{Sms}, \texttt{SMS} \\
\cmidrule(lr){2-3}

& Malware-related\newline terminology
& \texttt{rojan}, \texttt{ojan}, \texttt{roja}, \texttt{hish}, \texttt{jacke}, \texttt{jacker} \\
\cmidrule(lr){2-3}

& Escaped or\newline encoded text
& \texttt{\textbackslash u0}, \texttt{\textbackslash u17}, \texttt{u17} \\
\cmidrule(lr){2-3}

& Code/package fragments
& \texttt{com.ex}, \texttt{m.exa}, \texttt{om.exa}, \texttt{code>} \\

\midrule

\multirow{3}{*}[-5ex]{Benign}
& \textbf{Documentation or \newline code formatting}
& \texttt{ref}, \texttt{\#\#\#}, \texttt{//}, \texttt{if}, \newline \texttt{(space)*(space)} \\
\cmidrule(lr){2-3}

& Android resource\newline references
& \texttt{R.str}, \texttt{R.stri}, \texttt{R.st}, \texttt{.stri}, \texttt{.strin} \\
\cmidrule(lr){2-3}

& UI/application\newline development terms
& \texttt{lutt}, \texttt{lutte}, \texttt{lutter}, \texttt{olor}, \texttt{olo}, \texttt{lor}, \texttt{View} \\

\bottomrule
\end{tabular}
\end{table}

%% file: Text/7_Discussion_v4.tex
\section{Discussion}
\label{sec:discussion}

\paragraph{README-Based Malware Collection and Classification}
An important finding from our automated pipeline for collecting malware source code is that README files not only document developers’ intentions and project functionality but also serve as useful signals for malware classification at scale.
We use a classic combination of TF-IDF and LinearSVC to demonstrate that README-based malware retrieval is feasible and to establish a baseline for this task. 
The one-week deployment study reveals a gap between local and real-world performance, suggesting that README files in the real world are more complex than those in the local test set.
In particular, multilingual documents, short descriptions, and READMEs that contain rich structural information but limited textual content pose greater challenges for classification.
In this context, the confidence score threshold primarily serves as a quality-control mechanism to filter uncertain cases in real deployments, rather than as a parameter used solely to optimize classification performance.

In future work, metadata collected and parsed from malware README files may support further security analysis.
For example, researchers could extend the scope of collection to developer platforms other than GitHub to identify and collect more malware projects.
Another possible direction is to extract functionality, targets, and techniques from README documents to enable more specific studies.
For instance, researchers could collect malware projects whose READMEs claim the ability to detect or evade virtual machines, and then analyze their implementations to improve the design of security mechanisms for virtualized environments.

\paragraph{Lifecycle-Level Malware Analysis}
The value of the MASCOT-Android dataset lies not only in supporting malware evolution analysis and detection research.
But also in providing source-level malware representations that can be linked to subsequent features generated through compilation, decompilation, and static and dynamic analysis.
Such datasets will provide a complete lifecycle view of malware, enabling researchers to examine malware features and their transformations across stages.
For example, when analyzing different obfuscation methods, relying solely on decompiled code may not accurately capture the impact of obfuscation.
In contrast, combining source code with decompiled code allows researchers to examine how different obfuscation techniques interfere with decompilers and how such interference further affects subsequent static and dynamic feature analysis.

\paragraph{Code-reuse Network Annotated With LLM Assistance}
Section~\ref{sec:code_reuse_LLM} presents an overview of the extent to which Android malware source code specimens exhibit evidence of LLM assistance.
We observe that traces of LLM assistance in Android malware generally correspond to several milestones in the development of LLMs within the code reuse network.
Although we identify a representative example consistent with this trend, we should not conclude that a single malware specimen was directly assisted by LLMs, given the current limitations of code-oriented LLM detectors and other reasons~\cite{jiang2026scribe}.
Nevertheless, the code-reuse network annotated with LLM assistance provides a suitable basis for analyzing how LLM signals propagate during malware development.

\paragraph{Symbolic Information Ablation Analysis in Malware Source Code}
The symbolic information ablation analysis in Section~\ref{sec:ablation_experiments} reveals a clear distinction between the semantic utility and discriminative utility of symbolic information in the malware source code.
Comments, class names, function names, and variable names can help researchers interpret malicious behavior and improve the readability of decompiled code, but they contribute little to classification in our experimental settings.
In comparison, import statements expose library dependencies and provide more discriminative cues for malware detection.
Researchers should select symbolic information according to their research objectives and the target task.

\paragraph{Limitations}
First, in the described MASCOT-Android dataset, all malicious source code specimens are collected from GitHub. 
Although GitHub provides development histories and rich metadata that can support future research, relying on a single platform may introduce bias because attackers use multiple channels to distribute malware. 
Second, due to the lack of an LLM detector specifically designed for source code recently, the visualization in section~\ref{sec:code_reuse_LLM} may reflect only overall trends rather than reliable specimen-level evidence.
Finally, while preprocessing README files, we preserved some structural information using special characters, such as square brackets, to indicate a figure. 
This method cannot preserve the actual content represented by these structures, such as figures. 
Therefore, future research could explore methods to extract accurate structural information and heterogeneous content from malware-related repositories.

\textbf{Data availability.} The disarmed MASCOT-Android, automated collection pipeline, and other tools are available on GitHub: \url{https://github.com/BojingLi24/MASCOT_Android}.

%% file: Text/8_Conclusion_v4.tex
\section{Conclusion}
\label{sec:conclusion}

In this paper, we present MASCOT-Android, a curated dataset of Android malware source code, along with an automated pipeline for collecting malware source code.
The MASCOT-Android dataset contains 1,093 human-reviewed Android malware source code samples and rich metadata.
The proposed pipeline searches GitHub for candidate malware source code repositories, identifies relevant projects, and collects source code specimens, achieving an average accuracy of 96.28\% and an FPR of 1.06\% on local evaluations.
In real-world collection experiments, thresholding confidence scores effectively filters out low-confidence predictions in complex scenarios, demonstrating its practical utility.

We further conduct two case studies to demonstrate the unique value of malware source code for the security community.
The first case study visualizes evolutionary relationships among malware specimens and provides traces that attackers use LLMs to develop malware and that LLM-assisted code propagates through code-reuse networks.
The second case study evaluates the classification utility of symbolic information available in malware source code, including import statements, comments, class names, function names, and variable names.
Overall, we hope that the proposed MASCOT-Android dataset and collection pipeline will support malware analysis by providing reliable source-level artifacts that preserve unobfuscated and untransformed features.

\begin{acks}
We thank GPTZero for providing API access, which was used as a comprehensive and robust LLM trace detector in Section~\ref{sec:code_reuse_LLM}. 
\end{acks}